\documentclass[aps,prb,superscriptaddress,twocolumn]{revtex4}
\usepackage{bm}
\usepackage{graphicx}

\newcommand{\dave}[1]{\langle\hspace{-0.9mm}\langle#1\rangle\hspace{-0.9mm}\rangle}
\newcommand{\bigdave}[1]{\big\langle\hspace{-1.1mm}\big\langle#1\big\rangle\hspace{-1.1mm}\big\rangle}
\newcommand{\Bigdave}[1]{\Big\langle\hspace{-1.5mm}\Big\langle#1\Big\rangle\hspace{-1.5mm}\Big\rangle}
\newcommand{\biggdave}[1]{\bigg\langle\hspace{-1.95mm}\bigg\langle#1\bigg\rangle\hspace{-1.95mm}\bigg\rangle}

\begin{document}

\title{Path integral Monte Carlo for dissipative many-body systems}

\author{Luca~Capriotti}
\affiliation{Istituto Nazionale per la Fisica della Materia (INFM),
  Unit\`a di Ricerca di Firenze,
  via G. Sansone 1, I-50019 Sesto Fiorentino (FI), Italy.}
\affiliation{Kavly Institute for Theoretical Physics
  and Department of Physics, University of California,
  Santa Barbara CA 93106-4030.}

\author{Alessandro~Cuccoli}
\affiliation{Dipartimento di Fisica dell'Universit\`a di Firenze,
  via G. Sansone 1, I-50019 Sesto Fiorentino (FI), Italy.}
\affiliation{Istituto Nazionale per la Fisica della Materia (INFM),
  Unit\`a di Ricerca di Firenze,
  via G. Sansone 1, I-50019 Sesto Fiorentino (FI), Italy.}

\author{Andrea~Fubini}
\affiliation{Dipartimento di Fisica dell'Universit\`a di Firenze,
  via G. Sansone 1, I-50019 Sesto Fiorentino (FI), Italy.}
\affiliation{Istituto Nazionale per la Fisica della Materia (INFM),
  Unit\`a di Ricerca di Firenze,
  via G. Sansone 1, I-50019 Sesto Fiorentino (FI), Italy.}

\author{Valerio~Tognetti}
\affiliation{Dipartimento di Fisica dell'Universit\`a di Firenze,
  via G. Sansone 1, I-50019 Sesto Fiorentino (FI), Italy.}
\affiliation{Istituto Nazionale per la Fisica della Materia (INFM),
  Unit\`a di Ricerca di Firenze,
  via G. Sansone 1, I-50019 Sesto Fiorentino (FI), Italy.}

\author{Ruggero~Vaia}
\affiliation{Istituto di Fisica Applicata `N. Carrara'
  del Consiglio Nazionale delle Ricerche,
  via Panciatichi~56/30, I-50127 Firenze, Italy.}
\affiliation{Istituto Nazionale per la Fisica della Materia (INFM),
  Unit\`a di Ricerca di Firenze,
  via G. Sansone 1, I-50019 Sesto Fiorentino (FI), Italy.}

\date{\today}

\begin{abstract}
We address the possibility of performing numerical Monte Carlo
simulations for the thermodynamics of quantum dissipative systems.
Dissipation is considered within the Caldeira-Leggett formulation,
which describes the system in the path-integral formalism through the
inclusion of an influence action that is bilocal and quadratic in the
system's coordinates. At a first sight the usual direct approach of
discretizing the path integral could seem feasible, but complications
arise when one tries to introduce a physically meaningful dissipation
kernel: in particular its imaginary-time dependence turns out to be
severely singular and difficult to evaluate analytically, in spite of
the simple expressions for its Matsubara components. We therefore
propose to face the numerical problem using Fourier path-integral Monte
Carlo, that can be formulated in two different ways: transforming the
continuous paths and then truncating the high Fourier components (with
possible improvements upon the truncation procedure), or performing the
Fourier transformation upon the discretized paths. The latter choice
leads to a simpler formulation and allows for a better control of the
extrapolation to the limit of infinite Trotter number. The method is
implemented for a single nonlinear particle with Ohmic dissipation and
for a $\phi^4$ chain with Drude-like dissipation.
\end{abstract}

\maketitle

\section{Introduction}

In the last decades the interest in quantum dissipation~\cite{Weiss99}
has come mainly from the study of mesoscopic systems, which have been
experimentally fabricated and theoretically analyzed. In such systems,
the characteristic quantum effects involve a macroscopic number of
particles. The sizeable dimension of the devices implies that the
relevant dynamical variables can couple to a very large number of
degrees of freedom of the surrounding environment (or dissipation
bath): this coupling can be described macroscopically without caring
for the details of the interaction, and can result in dramatic changes
in the behavior of the system. For instance, the dissipative phase
transition in Josephson-junction arrays~\cite{TakahideEA2000} (JJA).

While the classical thermodynamics is unaffected by dissipation, its
quantum counterpart is substantially modified, and it constitutes
therefore an ideal field to study the genuine interplay between quantum
fluctuations and dissipation, which leads in general to interesting
physics in the regimes of high quantum coupling and/or low temperature.

The issue of evaluating thermodynamic quantities in a
quantum-dissipative system was recently faced by an extension of the
effective-potential method~\cite{CRTV97,CFTV99}, that is very fruitful
in the regime of intermediate quantum coupling. However, a more
powerful tool is required when the aim is to study dramatic effects,
as, for instance, the dissipative phase transition from superconducting
to insulating behavior in JJA predicted by mean-field theory.
Unfortunately, a suitable theoretical approach, allowing a faithful
comparison with the experimental findings in the regime of high quantum
coupling, is still lacking.

In this paper, we discuss an efficient path-integral Monte Carlo (PIMC)
approach can be implemented. In Section~\ref{s.dissipative} the basic
formalism and the connection with the phenomenological description
dissipation are reviewed. The customary approach to Monte Carlo is set
up in Section~\ref{s.RPIMC}, where some difficulties are pointed out;
this leads us to consider Fourier PIMC~\ref{s.FPIMCc} basically
extending the standard approach developed by many authors in the 80ies,
involving the transformation to Matsubara components and their
truncation by partial averaging, that by the way leads to a
reformulation of the effective potential method. We propose a slightly
different scheme for the numerical computation framework in
Section~\ref{s.FPIMCd}, which overcomes some ambiguities of the former.
Eventually, in Section~\ref{s.FPIMCd_appl} the latter method is applied
for two reference models: it appears that working with Fourier
transformed variables, possibly using the knowledge of the exact
quantum harmonic propagator, gives reliable results for many-body
systems with reasonable numerical effort.

\section{Path-integral for the dissipative system}
\label{s.dissipative}

\subsection{Formalism}

In this paper we consider the study of dissipation effects onto the
thermodynamics of a quantum system with Hamiltonian
\begin{equation}
 \hat{\cal H}=\frac{\hat p^2}{2m}+V(\hat q)~.
\end{equation}

The {\it Caldeira-Leggett} (CL) \cite{CaldeiraL81,CaldeiraL83a} model
considers the system of interest as linearly interacting with a bath of
harmonic oscillators, whose coordinates can be integrated out from the
path integral, leaving the CL euclidean action:
\begin{eqnarray}
 S[q] = \int_0^{\beta\hbar}\frac{du}\hbar
 \left[ \frac m2\, \dot q^2(u) + V\Big(q(u)\Big) \right]
 + S^{\rm{(nl)}}[q] ~~~~~~&&
\label{e.S}
\\
 S^{\rm{(nl)}}[q] = - \frac{m}{4\hbar}\int_0^{\beta\hbar}\!\!\! du
 \int_0^{\beta\hbar}\!\!\! du'\,k(u{-}u')\,\Big[q(u){-}q(u')\Big]^2.~&&
 \label{e.Snl}
\end{eqnarray}
The kernel $k(u)$ depends on the temperature $T=\beta^{-1}$ and is a
symmetric and periodic function of the imaginary-time $u$,
$k(u)=k(-u)=k(\beta\hbar-u)$; its functional form depends on the
spectral density of the environmental bath~\cite{Weiss99}; moreover, it
has a vanishing average, $\int_0^{\beta\hbar}du\,k(u)=0$. Thanks to the
last property, one can write the nonlocal dissipative action also as
\begin{equation}
 S^{\rm{(nl)}}[q] = \frac{m}{2\hbar}\int_0^{\beta\hbar}\!\!\! du
 \int_0^{\beta\hbar}\!\!\! du'\,k(u{-}u')~q(u)\,q(u')~.
\end{equation}

The density matrix elements in the coordinate representation are
expressed by Feynman's path integral as
\begin{equation}
 \rho(q'',q')= \int_{q'}^{q''}{\cal D}[q]~e^{-S[q]}~,
\label{e.rhoqq}
\end{equation}
where the path integration is defined as a sum over all paths
$q(u)$, with $u\in[0,\beta\hbar]$, $q(0)=q'$ and
$q(\beta\hbar)=q''$, and the partition function reads
\begin{equation}
 {\cal Z}=\oint{\cal D}[q]~e^{-S[q]}~.
\label{e.Z}
\end{equation}

The usual procedure for the phenomenological identification of $k(u)$
consists in comparing its explicit expression (in terms of the
dynamical variables of the oscillator bath) with the analogous
expression of the (retarded) damping function $\gamma(t)$ one gets in
deriving the (classical or quantum) Langevin equation of motion from
the same composite Hamiltonian~\cite{Weiss99},
\begin{equation}
 m\,\ddot q + m \int dt'\, \gamma(t-t')\,\dot q(t') + V'(q) = f(t) ~,
\label{e.langevin}
\end{equation}
where $f(t)$ is the fluctuating force. The relation that is found
between $k(u)$ and $\gamma(t)$ can be expressed in a simple way as a
relation between their respective Matsubara transform,
\begin{equation}
 k_n= \int_0^{\beta\hbar} du ~ e^{-i\nu_n u}~k(u)~,
 ~~~~~~ \nu_n=\frac{2\pi n}{\beta\hbar}~,
\end{equation}
and Laplace transform,
\begin{equation}
 \gamma(z)=\int_0^\infty dt~e^{-zt}~\gamma(t) ~,
\end{equation}
and reads
\begin{equation}
 k_n~=~|\nu_n|~\gamma\big(z{=}|\nu_n|\big) ~.
\label{e.kn}
\end{equation}
Here it is apparent that $k_0=0$, i.e. the `local part' is assumed to
be fully included as a quadratic term in the potential. The following
completeness/orthogonality relations have to be taken into account:
\begin{equation}
 \sum_{n=-\infty}^\infty e^{i\nu_nu} = \beta\hbar~\delta(u) ~,
 \label{e.compl1}
\end{equation}
where $\delta(u)=\delta(u+\beta\hbar)$ is the periodic delta function,
and its inverse
\begin{equation}
 \int_0^{\beta\hbar} du~ e^{i\nu_nu}
 = \frac{e^{i\nu_n\beta\hbar}-1}{i\nu_n}
 = \beta\hbar~\delta_{n0} ~.
 \label{e.compl2}
\end{equation}

\subsection{Ohmic and Drude dissipation}
\label{ss.OhmDru}

In the above dynamical equation (\ref{e.langevin}) the bath spectral
density is assumed to be such to reproduce the most useful
phenomenological models, namely:
\begin{itemize}
\item[{\em i)}] {\it Ohmic (or Markovian) dissipation}. This is
characterized by the absence of memory in the dissipative term, and
corresponds to assuming a separation of time scales: the time scale
with which the bath responds to changes in the system is much smaller
than the system's typical times. In this case dissipation can be
described by one constant parameter, $\gamma$:
\begin{equation}
 \gamma(t)=\gamma~\delta(t-0^+) ~,
 ~~~~~~ \gamma(z)=\gamma ~.
\label{e.ohmic}
\end{equation}
\item[{\em ii)}] {\it Drude-like dissipation}. Here the bath responds
on a time scale $\omega_{{}_D}^{-1}$ which is comparable to the
system's typical times:
\begin{equation}
 \gamma(t)=\gamma\,\omega_{{}_D}\,e^{-\omega_{{}_D}\,t} ~,
 ~~~~~~ \gamma(z)= \gamma \frac{\omega_{{}_D}}{\omega_{{}_D}+z} ~.
\label{e.drude}
\end{equation}
Therefore, there are two parameters which describe dissipation,
the intensity $\gamma$ and the response frequency (or `spectral
width') $\omega_{{}_D}$; for $\omega_{{}_D}\to\infty$, i.e., fast
bath response, the Ohmic form is recovered.
\end{itemize}
Note that $\gamma(z)$ has the dimension of a frequency, while $k_n$ is
a squared frequency. Only the above two cases will be considered in
what follows; although, of course, the actual physics of a problem
could give more appropriate definitions of $\gamma(t)$.

\subsection{Imaginary-time kernel for Ohmic dissipation}

From the above formulas it follows that the relation connecting
the imaginary-time kernel $k(u)$ with the (assumed known) Laplace
transform $\gamma(z)$ of the damping function $\gamma(t)$ is
\begin{equation}
 k(u) = \frac1{\beta\hbar} \sum\limits_{n=-\infty}^{\infty}
 e^{i\nu_n u}~|\nu_n|~\gamma\big(z=|\nu_n|\big)~.
\end{equation}
The point is that the simple cases above give rather involute
results for $k(u)$. Let us calculate it in the Ohmic
case~(\ref{e.ohmic}), invoking a criterion of mean convergence for
the resummation:
\begin{equation}
 \frac{\beta\hbar}\gamma~\tilde k(u)
 = \sum\limits_{n=-\infty}^{\infty} ~|\nu_n|~e^{i\nu_n u}
 = -\frac{\pi}{\beta\hbar}
 \bigg(\sin\frac{\pi\,u}{\beta\hbar}\,\bigg)^{-2}~.
\end{equation}
While it might be useful to note that
\begin{equation}
 -\frac{\pi}{\beta\hbar}
 \bigg(\sin\frac{\pi\,u}{\beta\hbar}\,\bigg)^{-2}
 = \partial_u\cot\frac{\pi{u}}{\beta\hbar}
 = \frac{\beta\hbar}{\pi}~\partial_u^2~ \ln\sin\frac{\pi{u}}{\beta\hbar} ~,
\end{equation}
so that one has alternative expressions,
\begin{equation}
 \tilde k(u)
 = \frac\gamma{\beta\hbar}~ \partial_u\cot\frac{\pi{u}}{\beta\hbar}
 = \frac\gamma\pi~ \partial^2_u \ln\sin\frac{\pi\,u}{\beta\hbar} ~,
\end{equation}
one can see that the requirement $k_0=\int{du}~k(u)=0$ is not satisfied
and that to fulfil it one must subtract from the expression found --
this is the reason why we used the tilde in the notation $\tilde k(u)$
-- the product of a (periodic) delta function $\delta(u)$ by an
infinite constant:
\begin{equation}
 k(u) = \tilde k(u) - \tilde k_0\,\delta(u)
\end{equation}
where
\begin{equation}
 \tilde k_0 =
 \int\limits_\epsilon^{\beta\hbar-\epsilon} du~ \tilde k(u)
 = -\frac{2\gamma}{\beta\hbar}~\cot\frac{\pi\epsilon}{\beta\hbar}
 ~~~\mathop{\longrightarrow}\limits_{\epsilon\to{0}}~~-\infty ~;
\end{equation}
one can indeed verify that the correct Matsubara transform
$|\nu_n|\gamma$ is obtained thanks to the cancelation of two
divergences,
\begin{eqnarray}
 k_n &=&\int_0^{\beta\hbar} du~ \tilde k(u)~(e^{-i\nu_nu}-1)
 \nonumber\\
 &=& \frac\gamma{\beta\hbar} \int_0^{\beta\hbar} du
 ~\partial_u \cot\frac{\pi{u}}{\beta\hbar}~(e^{-i\nu_nu}-1)
\nonumber\\
 &=& \frac\gamma{\beta\hbar}~ \bigg[\cot\frac{\pi{u}}{\beta\hbar}
 ~(e^{-i\nu_nu}{-}1)\bigg]_\epsilon^{\beta\hbar-\epsilon}
 \nonumber\\
 &&~~~~ + \frac{i\gamma\nu_n}{\beta\hbar}~
 \int_0^{\beta\hbar}du
 \,\cot\frac{\pi{u}}{\beta\hbar}~e^{-i\nu_nu}
\nonumber\\
 &=& \frac{2\gamma}{\beta\hbar}~\cot\frac{\pi\epsilon}{\beta\hbar}
 \bigg(1-\cos\frac{\pi\epsilon}{\beta\hbar}\bigg)
 \nonumber\\
 &&~~~~ + i\gamma\nu_n~ \int_0^{\pi}\frac{dx}\pi
 \,\cot x~e^{-2inx}
\nonumber\\
 &=& O(\epsilon) + i\gamma\nu_n~ (-i\,{\rm{sign}}\,n)
 = \gamma~ |\nu_n| ~.
\end{eqnarray}

\section{Real-space PIMC}
\label{s.RPIMC}

The standard PIMC approach consists in approximating the partition
function~(\ref{e.Z}) by discretizing the paths $q(u)$ on a finite mesh.
Namely, the imaginary-time interval $[0,\beta\hbar]$ is divided into
$P$ slices of finite duration $\varepsilon=\beta\hbar/P$, $P$ being the
so called {\em Trotter number}. Each whole path
$\big\{q(u),~u\in[0,\beta\hbar]\big\}$ turns into the $P$ discrete
quantities $\big\{q_\ell=q(\ell\varepsilon)\big\}$, with the
periodicity condition $q_0\equiv{q_P}$, and the action becomes:
\begin{eqnarray}
 S_P = \sum_{\ell=1}^P \bigg[ \frac{mP}{2\beta\hbar^2}~
 (q_\ell-q_{\ell-1})^2
 + \frac\beta P~ V(q_\ell) \bigg] + S^{\rm{(nl)}}_P ~~~&&
\label{e.SP}
\\
 S^{\rm{(nl)}}_P = - \frac{m\beta^2\hbar}{4P^2} \sum_{\ell,\ell'=1}^P
 k_{\ell-\ell'}~(q_\ell-q_{\ell'})^2~.~~~&&
 \label{e.SPnl}
\end{eqnarray}
The partition function is approximated by
\begin{equation}
 {\cal Z}_P= \left(\frac{mP}{2\pi\hbar^2\beta}\right)^{P/2}
 \int \prod_{\ell=1}^P dq_\ell~~e^{-S_P}~.
\label{e.RZP}
\end{equation}
In the standard PIMC procedure the thermodynamic averages (say, $G_P$)
obtained from this multiple integral are evaluated by a stochastic
simulation, e.g., the Metropolis algorithm for configuration sampling;
this is to be done for large enough values of $P$, and the exact result
$G=G_\infty$ is estimated by extrapolating~\cite{TakahashiI1984} the
calculated values $G_P$. For the discrete kernel $k_\ell$ that
approximates the singular function $k(u)$, it is reasonable to keep a
piecewise approximation, namely, for $\ell\neq{0}$
\begin{equation}
 k_\ell = \frac1\varepsilon\!\!\!
 \int\limits_{\varepsilon(\ell-\frac12)}^{\varepsilon(\ell+\frac12)}
 \!\!\!\!\!\! du~\tilde k(u)
 =\frac{\gamma{P}}{(\beta\hbar)^2}\bigg[
 \cot\frac\pi{P}\big(\ell{+}{\textstyle\frac12}\big)
 - \cot\frac\pi{P}\big(\ell{-}{\textstyle\frac12}\big) \bigg];
\label{e.klne0}
\end{equation}
for large $P$ one has
\begin{equation}
 k_\ell \simeq -\frac{\pi\gamma}{(\beta\hbar)^2}
 ~\bigg(\sin\frac{\pi\ell}P\,\bigg)^{-2}
 ~~\mathop{\sim}_{\ell\ll{P}}~~ \frac{P^2}{\ell^2}~,
\label{e.klne0a}
\end{equation}
and for $\ell={0}$
\begin{eqnarray}
 k_0 &=& \frac1\varepsilon \int_{-\varepsilon/2}^{\varepsilon/2}du~k(u)
 = \frac1\varepsilon \bigg[\int_{-\varepsilon/2}^{\varepsilon/2}
 - \int_0^{\beta\hbar} \bigg]~du~\tilde k(u)
\nonumber\\
 &=& -\frac1\varepsilon \int_{\varepsilon/2}^{\beta\hbar-\varepsilon/2}
 du~\tilde k(u) = \frac{2\gamma{P}}{(\beta\hbar)^2}~\cot\frac{\pi}{2P} .
\label{e.kleq0}
\end{eqnarray}
The choice~(\ref{e.klne0}) should be preferred to~(\ref{e.klne0a})
since it ensures the exact vanishing of $\sum_\ell~k_\ell$. However, it
is apparent that $k_0$ does not contribute to the
action~(\ref{e.SPnl}).

The interaction along the Trotter direction involves all pairs (which
is very bad from the point of view of the code efficiency) although it
is rapidly decreasing ($\sim\ell^{-2}$). This suggest the possibility
of cutting the interaction beyond, say, $R^{\rm{th}}$ neighbors
(keeping only $|\ell-\ell'|<R$); a rough calculation can be made
assuming that the kinetic term dominates, i.e. that
$(x_\ell{-}x_{\ell-1})^2\sim{g^2/(c^2tP)}$, which gives a ratio between
the discarded and the included dissipative interaction energy
$\sim{1/\ln{R}}$. In any case, it turns out that a simulation along
these lines requires to deal with long-ranged summations whose
short-range part is highly singular; moreover, if one would like to
consider more physical dissipation kernels, e.g., the Drude one, the
calculation of $k(u)$ and of $k_\ell$ becomes very involute in spite of
the simple expression of $k_n$.

\section{Fourier PIMC with continuous imaginary time}
\label{s.FPIMCc}

In order to overcome the above mentioned difficulties, let us try now
to face the problem from another point of view: since we know as
`initial input' the Matsubara components of the kernel, $k_n$, it is
worth to explore the possibility of using the simulation technique
based on the sampling of Fourier components of the path $q(u)$. We will
follow the scheme of Refs.~\onlinecite{Doll84,DollCF85,TopperT92,%
EleftheriouDCF99,NeirottiFD00,MielkeST00}, with some
modifications~\cite{LobaughV1992} that seem to improve upon their
approach when the so called {\it partial averaging} is performed.

The Fourier transform of the closed path $q(u)$, $u\in[0,\beta\hbar]$,
$q(0)=q(\beta\hbar)$, reads:
\begin{eqnarray}
 q(u)&=&\sum_{n=-\infty}^\infty q_n~e^{-i\nu_nu}
 \equiv \bar q + \sum_{n=1}^\infty q_n(u)
\nonumber\\
 q_n(u)&\equiv& 2 \big( x_n \cos\nu_nu + y_n \sin\nu_nu \big)~,
\label{e.FT}
\end{eqnarray}
where $q_n\equiv{x_n+iy_n}=q_{-n}^*$ since $q(u)$ is real, so that
$x_n=x_{-n}$ and $y_n=-y_{-n}$. Using the completeness and
orthogonality relations, Eqs.~(\ref{e.compl1}) and~(\ref{e.compl2}),
the inverse transform is found to be
\begin{equation}
 q_n=\int \frac{du}{\beta\hbar}~q(u)~e^{i\nu_nu}~,
\end{equation}
and obviously $q_0\equiv\bar{q}$ is the average point of the path.

In terms of the transformed variables the action (\ref{e.S}) takes the
form
\begin{equation}
 S[q] = \frac{\beta m}2 \sum_{n=-\infty}^\infty (\nu_n^2+k_n)~|q_n|^2
 + \int_0^{\beta\hbar} \frac{du}\hbar~ V\big(q(u)\big)~,
\label{e.FS}
\end{equation}
which accounts in a simple way for the nonlocal dissipative part,
at the price of leaving the integral involving the potential,
whose argument is to be meant as expressed as in Eq.~(\ref{e.FT}).
The path integral~(\ref{e.Z}) for the partition function
transforms into
\begin{eqnarray}
 {\cal Z} &=& C \int d\bar{q} \prod_{n=1}^\infty
 \bigg[\frac{\beta m\nu_n^2}{\pi}\int d^2q_n~
 e^{-\beta m (\nu_n^2+k_n)|q_n|^2}\bigg]
\nonumber\\
 &&\hspace{15mm} \times
 ~\exp\bigg\{- \int_0^{\beta\hbar} \frac{du}\hbar~
 V\big(q(u)\big)\bigg\}~,
\label{e.FZ}
\end{eqnarray}
where $|q_n|^2=x_n^2+y_n^2$ and $d^2q_n=dx_n\,dy_n$ and
\begin{equation}
 C = \sqrt{\frac{m}{2\pi\hbar^2\beta}}~.
\end{equation}
The measure can be easily checked in the free-particle
nondissipative limit. One can think this expression as the
Gaussian average of the last exponential:
\begin{equation}
 {\cal Z} = C~e^{-\beta\mu} \int d\bar{q}~
 \biggdave{\exp\bigg\{- \int_0^{\beta\hbar}
 \frac{du}\hbar~ V\big(q(u)\big)\bigg\} }~,
\label{e.FZ1}
\end{equation}
with
\begin{equation}
 \mu = \frac 1\beta\sum_{n=1}^\infty\, \ln\frac{\nu_n^2+k_n}{\nu_n^2}~,
\label{e.mu}
\end{equation}
and the nonvanishing moments
\begin{equation}
 \bigdave{x_n^2}=\bigdave{y_n^2}=
 \frac1{2\beta m}~\frac1{\nu_n^2+k_n}~,
\label{e.xyave}
\end{equation}
i.e., the $n$-th component of $q(u)$ has the variance
\begin{eqnarray}
 \bigdave{q_n^2(u)} &=&
 4\Big(\dave{x_n^2}\cos^2\nu_nu + \dave{y_n^2}\sin^2\nu_nu \Big)
 \equiv \alpha_n
\\
 \alpha_n &=& \frac2{\beta m}~\frac1{\nu_n^2+k_n}~.
\end{eqnarray}
A MC simulation based on Eq.~(\ref{e.FZ1}) involves a Metropolis
dynamics for the Fourier coefficients $\bar{q}$, $x_n$, and $y_n$, with
a truncation of the series (\ref{e.FT}), say, at $n=P$; this should
correspond to a standard simulation with Trotter number $P$.

On the other hand, the authors of of Refs.~\onlinecite{Doll84,DollCF85,%
TopperT92,EleftheriouDCF99,NeirottiFD00,MielkeST00} always expand
$q(u)$ after subtracting the initial point $q\equiv{q(0)}$, in a
$\sin$-only series, i.e.,
\begin{equation}
 q(u)= q + \sum_{n=1}^\infty a_n \sin \frac{\pi n u}{\beta\hbar}~.
\end{equation}
The difference of this choice resembles the one between the use of
fixed boundary conditions (stationary waves, nonuniform amplitude)
instead of periodic boundary conditions (plane waves, uniform
amplitude).

\subsection{Partial averaging}

The {\it partial averaging}~\cite{DollCF85} improves upon the rude
truncation of the Fourier series for $q(u)$, and basically relies upon
the Jensen inequality~\cite{Feynman1972}. Look again at
Eqs.~(\ref{e.FZ}) and~(\ref{e.FZ1}): these can be expressed as a
superposition of uncorrelated Gaussian averages $\langle
F(\{x_n,y_n\})\rangle_n$ upon the variables $x_n$ and $y_n$, and for
anyone of these averages we can choose to approximate
\begin{equation}
 \langle e^F \rangle_n \gtrsim e^{\langle F\rangle_n}~.
\end{equation}
Therefore, choosing to retain (and simulate) the Fourier
components up to $n=P$, one can estimate what is left over in the
exact average; separating the components that we want to keep (up
to $n=P$) from those which are to be averaged out, i.e.,
\begin{equation}
 q(u)=q_{{}_P}(u)+\xi_{{}_P}(u)
\end{equation}
with
\begin{eqnarray}
 q_{{}_P}(u)&=& \bar q + \sum_{n=1}^P q_n(u)~,
\label{e.qP}
\\
 \xi_{{}_P}(u) &=& \sum_{n=P+1}^\infty q_n(u)~,
\label{e.xiP}
\end{eqnarray}
one can immediately get
\begin{eqnarray}
 \bigdave{\xi_{{}_P}^2(u)} &=& \sum_{n=P+1}^\infty \Bigdave{q_n(u)}
 \equiv \alpha_{{}_P}
\nonumber\\
 \alpha_{{}_P}&=& \sum_{n=P+1}^\infty \alpha_n
 = \frac2{\beta m}\sum_{n=P+1}^\infty \frac1{\nu_n^2+k_n}~,
\end{eqnarray}
and apply the Jensen inequality for this part getting the
approximate (upper bound for the) partition function as a Gaussian
average $\dave{\cdots}_{{}_P}$ over the finite set of the first
$2P+1$ variables,
\begin{equation}
 {\cal Z} = C ~e^{-\beta\mu}\int d\bar{q}~
 \biggdave{\exp\bigg\{- \int_0^{\beta\hbar}
 \frac{du}\hbar~ V_{{}_P}\big(q_{{}_P}(u)\big)\bigg\} }_P~,
\label{e.FZ2}
\end{equation}
with an effective potential $V_{{}_P}$ given as the Gaussian
smearing $\dave{\cdots}_{\alpha_P}$ on the scale of
$\alpha_{{}_P}$,
\begin{equation}
 V_{{}_P}(q_{{}_P})=\Bigdave{V(q_{{}_P}+\xi_{{}_P})}_{\alpha_P}~,
\end{equation}
where $\dave{\xi^2_{{}_P}}_{\alpha_P}=\alpha_{{}_P}$. What makes this
result appealing compared to the previous approaches is the fact that
$\alpha_{{}_P}$ does not depend on $u$, as it occurs for the
'stationary wave' approach, so one can expect that even in the
nondissipative case this could be an improvement for PIMC coding.
Moreover, note that taking the roughest approximation, i.e. $P=0$, one
gets exactly the recipe for the effective potential introduced by
Feynman:
\begin{equation}
 {\cal Z}=C~e^{-\beta\mu} \int d\bar{q}~ e^{-\beta V_0(\bar{q})}~,
\end{equation}
where $V_0(\bar{q})=\bigdave{V(\bar{q}+\xi_0)}_{\alpha_0}$ is
broadened with
\begin{equation}
 \alpha_0=\frac2{\beta m}\sum_{n=1}^\infty \frac1{\nu_n^2+k_n}
 ~~\mathop{\longrightarrow}\limits_{k_n\to{0}}~~
 \frac{\beta\hbar^2}{12m}~,
\end{equation}
while $\mu\to{0}$ for $k_n\to{0}$.

\subsection{The variational effective potential}
\label{ss.Veff}

In view of improving the technique, one can speculate whether it is
possible to better account for the harmonic part, in the spirit of
Refs.~\onlinecite{GT85,GT86,FK86}. Let us first review how the improved
variational approximation arises in the present context. The aim is to
incorporate a frequency term in the Gaussian averages (\ref{e.xyave}),
i.e., in the variances appearing in Eq.~(\ref{e.FZ1}), and since there
is an overall integration over $\bar{q}$, the frequency
$\omega=\omega(\bar{q})$ can depend on it. Thus we rewrite
Eq.~(\ref{e.FZ}) as follows
\begin{eqnarray}
 {\cal Z} &=& C \int d\bar{q} \prod_{n=1}^\infty
 \bigg[\frac{\beta m\nu_n^2}{\pi}\int d^2q_n~
 e^{-\beta m \big[\nu_n^2+k_n+\omega^2\big]|q_n|^2}\bigg]
\nonumber\\
 & & \hspace{-3mm} \times~
 \exp\bigg\{\!\!- \int_0^{\beta\hbar} \! \frac{du}\hbar
 ~\Big[V\big(q(u)\big)
 - \frac m2\,\omega^2\,\big(q(u){-}\bar{q}\big)^2\Big] \bigg\}~,
\nonumber\\
 &\equiv& C \int d\bar{q}~ \biggdave{\exp\bigg\{- \int_0^{\beta\hbar}
 \frac{du}\hbar~ \delta V\big(q(u)\big)\bigg\} }~,
\label{e.FZom}
\end{eqnarray}
where
\begin{equation}
 \delta V\big(q(u)\big) \equiv V\big(q(u)\big)
 - \frac m2\,\omega^2\,\big(q(u){-}\bar{q}\big)^2 + \mu~,
\end{equation}
and now
\begin{eqnarray}
 \dave{x_n^2} &=& \dave{y_n^2}=
 \frac1{2\beta m}~\frac1{\nu_n^2+k_n+\omega^2}~,
\label{e.xyaveom}
\\
\nonumber\\
 \mu &=& \frac 1\beta\sum_{n=1}^\infty\,
 \ln\frac{\nu_n^2+k_n+\omega^2}{\nu_n^2}
 \mathop{\longrightarrow}\limits_{k_n\to{0}}~~
 \ln\frac{\sinh f}f~,
\label{e.muom}
\end{eqnarray}
with $f(\bar{q})=\beta\hbar\omega(\bar{q})/2$; note that the
integral over $\bar{q}$ always stays in front of the Gaussian
averages, so that any quantities `inside' it can naturally depend
on $\bar{q}$, and there is no need to emphasize this dependence.
Then, taking the Jensen approximation for all fluctuating
components, one gets
\begin{eqnarray}
 {\cal Z} &\gtrsim& C \int d\bar{q}~ e^{- \beta\,V_{\rm{eff}}(\bar{q})}
\label{e.ZVeff}
\\
 V_{\rm{eff}}(\bar{q})&=&
 \Bigdave{\delta V\big(q(u)\big) }
\nonumber\\
 &=& \dave{V(\bar{q}+\xi)}
 - \frac m2\,\omega^2(\bar{q})~\alpha_0(\bar{q}) +\mu(\bar{q})
\label{e.Veff}
\\
 \alpha_0(\bar{q}) &=& \frac 2{\beta m}\sum_{n=1}^\infty\,
 \frac1{\nu_n^2+k_n+\omega^2(\bar{q})}
\nonumber\\
 &&~~~\mathop{\longrightarrow}\limits_{k_n\to{0}}~~~
 \frac\hbar{2m\omega}\,
 \bigg(\coth f-\frac1f\bigg)~,
\label{e.alphaom}
\end{eqnarray}
Note that the dependence of $V_{\rm{eff}}$ on $u$ disappears upon
averaging. We have now to maximize the r.h.s. of Eq.~(\ref{e.ZVeff}),
i.e. to minimize the effective potential~(\ref{e.Veff}), in order to
determine $\omega^2(\bar{q})$. Since
\begin{equation}
 \frac{\partial\mu}{\partial\omega^2}=\frac m2\,\alpha_0 ~,
\end{equation}
a cancelation occurs and what is left is the known determination,
\begin{equation}
 \frac{\partial V_{\rm{eff}}}{\partial\omega^2} =
 \frac12 \Big(\dave{V''(\bar{q}+\xi)}-m\,\omega^2(\bar{q})\Big)
 \frac{\partial\alpha_0}{\partial\omega^2}=0 ~.
\end{equation}
This concludes the derivation of the effective potential. Note that
there is no need to introduce the parameter $w(\bar{q})$ of
Refs.~\onlinecite{GT85,GT86,FK86} and to optimize it.

\subsection{Improved partial averaging}

In order to retain the exact calculation of the first $P$ fluctuation
variables, let us split $q(u)$ as in Eqs.~(\ref{e.qP}-\ref{e.xiP}) and
introduce the frequency $\omega^2=\omega^2(q_0,...,q_{{}_P})$ in the
Gaussian variances we want to approximate, i.e., those labeled by
$n=P{+}1,...,\infty$~:
\begin{eqnarray}
 {\cal Z} &=& C \int d\bar{q} \prod_{n=1}^P
 \bigg[\frac{\beta m\nu_n^2}{\pi}\int d^2q_n~
 e^{-\beta m (\nu_n^2+k_n)|q_n|^2}\bigg]
\nonumber\\
 & &  \times \prod_{n=P+1}^\infty
 \bigg[\frac{\beta m\nu_n^2}{\pi}\int d^2q_n~
 e^{-\beta m \big[\nu_n^2+k_n+\omega^2\big]|q_n|^2}\bigg]
\nonumber\\
 & &  \times~
 \exp\bigg\{- \int_0^{\beta\hbar} \frac{du}\hbar~\Big[
 V\big(q(u)\big) -\frac m2\,\omega^2\,\xi_{{}_P}^2(u)
 \Big] \bigg\}~,
\nonumber\\
 &\equiv& C \int d\bar{q}~ \biggdave{\exp\bigg\{- \int_0^{\beta\hbar}
 \frac{du}\hbar~ \delta V\big(q(u)\big)\bigg\} }~,
\label{e.FZomP}
\end{eqnarray}
where
\begin{equation}
 \delta V\big(q(u)\big) = V\big(q_{{}_P}(u)+\xi_{{}_P}(u)\big)
 -\frac m2\,\omega^2\,\xi_{{}_P}^2(u)+\mu_{{}_P}~,
\end{equation}
with
\begin{equation}
 \mu_{{}_P}=\frac1\beta\bigg[ \sum_{n=1}^P \ln\frac{\nu_n^2+k_n}{\nu_n^2}
 +\sum_{P+1}^\infty \ln\frac{\nu_n^2+k_n+\omega^2}{\nu_n^2} \bigg]~.
\end{equation}
In order to perform the partial averaging, we take now the Jensen
approximation for the Gaussian components beyond the $P$-th one, so the
relevant variance is
\begin{equation}
 \alpha_{{}_P} = \bigdave{\xi_{{}_P}^2(u)}
 = \frac2{\beta m}\sum_{n=P+1}^\infty \frac1{\nu_n^2+k_n+\omega^2}~,
\label{e.alphaP}
\end{equation}
and the approximation reads
\begin{equation}
 {\cal Z} \gtrsim C \int d\bar{q}~
 \biggdave{\exp\bigg\{- \int_0^{\beta\hbar}
 \frac{du}\hbar~ V_{{}_P}\big(q_{{}_P}(u)\big)\bigg\} }_P~,
\end{equation}
with the effective potential
\begin{equation}
 V_{{}_P}(q) =
 \bigdave{V(q+\xi_{{}_P})}_{\alpha_P}
 -\frac m2\,\omega^2\,\alpha_{{}_P}+\mu_{{}_P}~,
\end{equation}
that actually depends on $\{q_0,...,q_{{}_P}\}$ since
$q=q_{{}_P}(u)$ is given by Eq.~(\ref{e.qP}). In order to optimize
$\omega$ we must minimize the {\it integral} of the effective
potential,
\begin{equation}
 0=\frac\partial{\partial\omega^2}~\int_0^{\beta\hbar}
 \frac{du}\hbar V_{{}_P}\big(q_{{}_P}(u)\big) ~,
\end{equation}
which, since
$\partial\mu_{{}_P}/\partial\omega^2=m\alpha_{{}_P}/2$, gives
\begin{equation}
 \int_0^{\beta\hbar} \frac{du}\hbar \bigg\{
 \frac12\Big[\bigdave{V''\big(q_{{}_P}(u)+\xi_{{}_P}\big)}_{\alpha_P}
 -m\,\omega^2\Big]\frac{\partial\alpha_{{}_P}}{\partial\omega^2}
 \bigg\}=0~,
\end{equation}
and definitely
\begin{equation}
 m\,\omega^2 = \int_0^{\beta\hbar} \frac{du}{\beta\hbar}
 ~ \bigdave{V''\big(q_{{}_P}(u)+\xi_{{}_P}\big)}_{\alpha_P}~.
\label{e.omP}
\end{equation}
The effective potential can therefore be written as
\begin{equation}
 V_{{}_P}(q) =
 \bigdave{V(q{+}\xi_{{}_P})}_{\alpha_P}
 -\frac {\alpha_{{}_P}}2\,
 \bigdave{V''(q{+}\xi_{{}_P})}_{\alpha_P} + \mu_{{}_P} ~,
\end{equation}
or, using the differential operator
$\Delta_{{}_P}=\frac12\,\alpha_{{}_P}\partial_q^2$, as
\begin{equation}
 V_{{}_P}(q) = (1-\Delta_{{}_P})\,e^{\Delta_P}~V(q) + \mu_{{}_P}
\end{equation}
Eqs.~(\ref{e.alphaP}) and~(\ref{e.omP}) are self-consistent for
any choice of the arguments $(q_0,...,q_{{}_P})$.

\subsection{Low-coupling approximation (LCA)}

In the above framework the frequency $\omega^2(q_0,...,q_{{}_P})$
depends on all simulated variables and the self-consistent
Eqs.~(\ref{e.alphaP}) and~(\ref{e.omP}) give rise to a
considerable complexity, even for one degree of freedom; indeed,
one should practically solve those equations after each MC move
except for very simple potentials as the quartic one just
discussed. Some kind of LCA is then necessary; among several
possibilities, the most reasonable choices for approximating
$\omega^2(q_0,...,q_{{}_P})$ with some $\omega_0^2$ are:

\begin{itemize}
\item[({\it i}\,)] leaving the only dependence on $q_0=\bar{q}$ by
averaging over the fluctuation coordinates $q_1,...,q_{{}_P}$, which
leaves the need of tabulating the resulting
$\omega_0^2(\bar{q})=\bigdave{\omega^2(q_0,...,q_{{}_P})}_P$;
therefore, choosing to insert $\omega$ also in the first $P$ Gaussian
averages in order to better describe the resulting probability
distribution, one has
\begin{equation}
 \omega_0^2(\bar{q}) = \bigdave{V''(\bar{q}+\xi_0)}
 = e^\Delta~V''(\bar{q}) ~,
\end{equation}
with $\Delta=\frac12\,\alpha\,\partial^2_q$ and the full pure-quantum
spread
\begin{equation}
 \alpha \equiv \alpha_0 = \bigdave{\xi_0^2}
 = \frac2{\beta m}\sum_{n=1}^\infty \frac1{\nu_n^2+k_n+\omega^2}~;
\end{equation}
\item[({\it ii}\,)] taking the above value in the minimum,
$\omega_0^2=\omega_0^2(\bar{q}{=}q_{\rm{m}})$ (of course, it is also
possible to take the improved LCA, i.e.the self-consistent HA (SCHA) of
$\omega_0^2(\bar{q})$), so that the above self-consistent equations are
solved only once.
\end{itemize}

\noindent The first choice reduces the complexity of the
self-consistent equations to the same one of the approach of
Refs.~\onlinecite{GT85,GT86,FK86}, and can then be used for problems
with few degrees of freedom, while the latter appears to be necessary
when facing many-body problems. In both cases the effective potential
has to be expanded in the same way. After splitting
\begin{equation}
 \omega^2(q_0,...,q_{{}_P})=\omega_0^2 + \delta\omega^2~,
\end{equation}
where (for simplicity the integral is omitted)
\begin{equation}
 \delta\omega^2=\frac1m\, e^{\Delta_P}V''\big(q_{{}_P}(u)\big)
 -\omega_0^2~,
\end{equation}
we use $\partial\mu_{{}_P}/\partial\omega^2=m\alpha_{{}_P}/2$ in
expanding
\begin{eqnarray}
 \mu_{{}_P} &\simeq& \mu_{{}_{0P}}+\frac m2\,\alpha_{{}_{0P}}
 \Big[ m^{-1}e^{\Delta_{{}_P}}V''\big(q_{{}_P}(u)\big)-\omega_0^2\Big]
\nonumber\\
 &\simeq& \mu_{{}_{0P}}
 + e^{\Delta_{{}_P}}\Delta_{{}_{0P}} V\big(q_{{}_P}(u)\big)
 - \frac m2\,\alpha_{{}_{0P}}\,\omega_0^2~,
\end{eqnarray}
where terms of order $\delta\omega^4$ are neglected, and replacing
this in the effective potential we get
\begin{eqnarray}
 V_{{}_P} (q) &\simeq&
 (1{-}\Delta_{{}_P}{+}\Delta_{{}_{0P}})\,e^{\Delta_{{}_P}} \,V(q)
 +\mu_{{}_{0P}}-\frac m2\, \alpha_{{}_{0P}}\omega_0^2
\nonumber\\
 &\simeq& (1{-}\delta\Delta_{{}_P})
 \,e^{\Delta_{0P}+\delta\Delta_{{}_P}} \, V(q)
 +\mu_{{}_{0P}}-\frac m2\, \alpha_{{}_{0P}}\omega_0^2~,
\nonumber
\end{eqnarray}
and, neglecting terms of order $\delta\Delta_{{}_P}^2$, the LCA
effective potential eventually reads
\begin{equation}
 V_{{}_P} (q) = e^{\Delta_{0P}}\, V(q)
 +\mu_{{}_{0P}}-\frac m2\, \alpha_{{}_{0P}}\omega_0^2 ~.
\end{equation}

Eventually, the expression for the partition function suitable for
numerical simulation reads
\begin{equation}
 {\cal Z} = C \int d\bar{q}~
 \biggdave{\exp\bigg\{- \int_0^{\beta\hbar}
 \frac{du}\hbar~ V_{{}_P}\big(q_{{}_P}(u)\big)\bigg\} }_P~.
\end{equation}
Other possibilities for a LCA are explored in
Ref.~\onlinecite{LobaughV1992}, where the above described approach was
also implemented for the Morse potential.

\section{Fourier PIMC with discrete imaginary time}
\label{s.FPIMCd}

In order to numerically evaluate the integral appearing in
Eq.~(\ref{e.Z}), we have seen in Section~\ref{s.RPIMC} that the
standard PIMC method divides the imaginary-time interval
$[0,\beta\hbar]$ into $P$ slices of width $\varepsilon=\beta\hbar/P$,
and that the coordinate $q(u)$ turns into the discrete quantities
$q_\ell=q(\ell\varepsilon)$. The partition function ${\cal{Z}}$ and the
other macroscopic thermodynamic quantities are obtained as the
$P\to\infty$ extrapolation of Eq.~(\ref{e.RZP}) and of the estimators
generated from it.

As mentioned in Section~\ref{s.RPIMC}, the application of this direct
PIMC approach to a dissipative system is made difficult by the fact
that the kernel $k(u{-}u')$ is explicitly known in terms of its
Matsubara transform $k_n$, i.e., Eq.~(\ref{e.kn}), rather than in the
imaginary-time domain~\cite{Weiss99}. In fact, it is given in terms of
the Laplace transform $\gamma(z)$ of the damping function $\gamma(t)$
appearing in the phenomenological Langevin equation~(\ref{e.langevin}),
and we have seen for Ohmic dissipation $\gamma(z)=\gamma$ that this
makes $k(u{-}u')$ long-ranged, while for a more realistic Drude
dissipation $k(u{-}u')$ becomes very hard to evaluate.

In the previous Section we realized that the Fourier path integral is
very convenient as far as the treatment of the dissipative nonlocal
action is concerned, because it enters the relevant expressions trough
the (assumed known) Matsubara components $k_n$. However, the continuous
imaginary-time approach used there has a general drawback (also present
in the nondissipative case) arising from the appearance of the integral
of the potential in the last exponent of Eq.~(\ref{e.FZ}).

The alternative we propose here is to start from the finite-$P$
expression~(\ref{e.RZP}) of the standard PIMC for the partition
function and make there a lattice ({\em discrete}) Fourier transform,
changing the integration variables from $q_\ell$ to $q_n$ by setting:
\begin{equation}
 q_\ell = \bar{q} + \sum_{n=1}^{P-1}~q_n~e^{i2\pi\ell n/P}~,
\end{equation}
so that:
\begin{eqnarray}
 {\cal Z}_P &=& \frac C{\beta^{P\over2}}\int\! d\bar q
 \int\prod_{\ell=1}^{P-1}dq_n
\nonumber\\
 && \times~
 \exp\Bigg\{\!-\frac{m\beta}2 \sum_{n=1}^{P-1}
 \big(\nu_{P,n}^2{+}k_n\big)\,|q_n|^2
\nonumber\\
 && \hspace{7mm}
 -\frac\beta P ~\sum_\ell V\bigg(\bar{q}{+}\sum_{n=1}^{P-1}q_n
 e^{i2\pi\ell n/P} \bigg)\Bigg\},
\label{e.ZP}
\end{eqnarray}
where $C$ is a temperature-independent normalization and $k_n$ is as
given in Eq.~(\ref{e.kn}). Comparing with the previous
expression~(\ref{e.FZ}), two significant differences appear: firstly,
the last term (integral of the potential along a path) is converted to
a well-defined summation that doesn't require further approximations;
secondly, the kinetic-energy term contains the finite-$P$ Matsubara
frequencies
\begin{equation}
 \nu_{P,n}\equiv\frac{2P}{\beta\hbar}\,\sin\frac{\pi n}P
\end{equation}
rather than $\nu_n=2\pi{n}/\beta\hbar$, which are approached for
$P{\to}\infty$. Thanks to these features the expression we got is {\em
exactly equivalent} to the standard finite-$P$
expression~(\ref{e.RZP}), a property that gives us control onto the
extrapolation of the results to $P{\to}\infty$.

Estimators for the relevant thermodynamic quantities can be obtained in
the usual ways~\cite{EleftheriouDCF99,NeirottiFD00}; for example, from
the thermodynamic relation $U=-\partial_\beta\ln{\cal{Z}}$, the
following estimator for the internal energy is found:
\begin{equation}
 U_P= V_P +{P\over 2\beta} - \sum_{n=1}^{P-1}
 \bigg[ \frac{2mP^2}{\beta^2\hbar^2}~\sin^2\frac{\pi n}P -
 {m}k_n \bigg]|q_n|^2~.
\end{equation}

For a given potential $V(\hat{q})$, it is convenient to devise a
characteristic energy scale $\epsilon$ (e.g., the barrier height for a
double well potential, the well depth for physical potentials that
vanish at infinity, etc.) and length scale $\sigma$ (such that
variations of $V$ comparable to $\epsilon$ occur on this length scale)
and write
\begin{equation}
 V(\hat{q})={\epsilon}\,v(\hat{q}/\sigma)~.
\end{equation}
In this way one better deals with the dimensionless coordinate
$\hat{x}=\hat{q}/\sigma$. If $x_{\rm{m}}$ is the absolute minimum of
$v(x)$, the harmonic approximation (HA) of the system is characterized
by the frequency $\omega_0$ given by
\begin{equation}
 \omega_0^2=\frac{\epsilon\,v''}{m\,\sigma^2}~,
 ~~~~~~ v'' \equiv v''(x_{\rm{m}}) ~;
\end{equation}
the coupling parameter $g$ for the system can be defined as the ratio
between the HA quantum energy-level splitting $\hbar\omega_0$ and the
overall energy scale $\epsilon$,
\begin{equation}
 g=\frac{\hbar\omega_0}{\epsilon}
 =\sqrt{\frac{\hbar^2 v''}{m\,\epsilon\,\sigma^2}}~.
\label{e.g}
\end{equation}
The case of weak (strong) quantum effects occurs when $g$ is small
(large) compared to 1. It is then easy to make use of dimensionless
variables only, {\it i.e.} to give energies in units of $\epsilon$,
lengths in units of $\sigma$, frequencies in units of $\omega_0$, and
so on; the reduced temperature is $t=1/(\epsilon\beta)$, the reduced
damping intensity is $\widetilde\gamma=\gamma/\omega_0$.

We can finally write a dimensionless expression for the partition
function~(\ref{e.ZP}) (for odd Trotter number $P=2N+1$):
\begin{eqnarray}
 {\cal Z}_P&=& C t^{P\over 2}\int\! d\bar x
 \int\! \prod_{n=1}^N da_ndb_n
\nonumber\\
 && \hspace{-4mm} \times~
 \exp\Bigg\{\! -\sum_{n=1}^N
 \bigg[\frac{4v''tP^2}{g^2}~\sin^2\frac{\pi n}P
 {+} \frac{v''}{t} K_n \bigg](a_n^2{+}b_n^2)
\nonumber\\
 &&  \hspace{30mm}
 -\frac{1}{tP}\sum_{\ell}v(x_\ell)\Bigg\}~,
\label{e.FZPo}
\end{eqnarray}
where $x_\ell=\bar{x}+2\sum_{n=1}^N [a_n\cos\frac{2\pi\ell{n}}P+
b_n\sin\frac{2\pi\ell{n}}P]$, $K_n=k_n/\omega_0^2$ and we have used the
symmetry properties of $k(u)$, so that $K_{P-n}=K_n$. The real Fourier
variables $\bar{x}$, $a_n$ and $b_n$ are dimensionless; the integrals
in Eq.~(\ref{e.FZPo}) may be numerically evaluated by standard Monte
Carlo sampling techniques, e.g., the Metropolis one.

\section{Fourier PIMC with discrete imaginary time: applications}
\label{s.FPIMCd_appl}

\subsection{Single particle in the double-well potential}

\begin{figure}
\includegraphics[bbllx=14mm,bblly=91mm,bburx=174mm,bbury=208mm,
 width=80mm,angle=0]{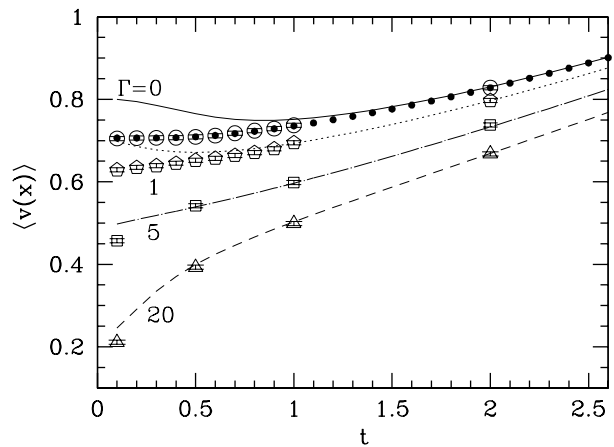}
\caption{\label{f.sp_q2}
Temperature dependence of the average potential energy
$\langle{v(x)}\rangle$ for the single particle in a quartic double
well, for $g=5$ and different values of the damping strength $\Gamma$.
Empty symbols are PIMC data, lines the predictions from the effective
potential method~\cite{CFTV99} and the filled circles are the exact
results for $\Gamma=0$. }
\end{figure}

As a first application we consider a particle in a quartic double well
potential $v(x)=(1-x^2)^2$ in presence of Ohmic dissipation, i.e.,
$k_n=2\pi(t/g)\Gamma n$, where $\Gamma$ is the damping strength in
units of $\omega_0$; the same model was already investigated in
Ref.~\onlinecite{CRTV97} by means of the effective-potential method
outlined in Section~\ref{ss.Veff}. In Fig.~\ref{f.sp_q2} we show the
Fourier PIMC results for the average potential energy
$\langle{v(x)}\rangle$ at the strong quantum coupling $g=5$, for
different values of the damping strength. The Monte Carlo data reported
in the figure represent the extrapolation to $P\to\infty$ of the
results obtained at $P=17$, 33, 65, and~129. First of all, for the
non-dissipative system ($\Gamma=0$) we observe the perfect agreement
between the exact results (obtained by numerical solution of the
Schr\"odinger equation) and the PIMC data, proving the reliability of
the PIMC code; for the dissipative model, the PIMC data provide a novel
reference to check the validity of the previous effective-potential
results~\cite{CFTV99}. In particular, the latter turns out to be
reliable at lower and lower temperature as the damping strength
increases: indeed, this is expected since the coordinate fluctuations
decrease with $\Gamma$, i.e., the coordinate-dependent quantities tend
to the classical behavior as an effect of dissipation.

\subsection{One-dimensional $\phi^4$ chain}

Let us now consider a many-body dissipative system, namely, the quantum
discrete $\phi^4$ chain, whose Hamiltonian may be written as
\begin{eqnarray}
 {\cal H}&=& \varepsilon_{{}_{\rm{K}}} \bigg[\,
 \frac{Q^2R}3\sum_{i=1}^M \hat p_i^2+V(\bm{\hat q})\, \bigg]~,
\label{e.Hphi4}
\\
 V(\bm{\hat q})&=& \frac{3}{2R}\sum_{i=1}^M\bigg[\, v(\hat q_i)
 +\frac{\,R^2}2\,(\hat q_i-\hat q_{i-1})^2\bigg] ~,
\label{e.Vphi4}
\end{eqnarray}
where $v(x)=(1-x^2)^2/8$, $Q$ is the quantum coupling and
$\varepsilon_{{}_{\rm{K}}}$ and $R$ are the kink energy and length,
respectively, in the classical continuum
limit~\cite{MakiT79,SchneiderS80,GTV88fi4}. In the above Hamiltonian
the number of particles in the chain is $M$ and periodic boundary
conditions are assumed. The canonical variables are such that
$[\hat{q}_i,\hat{p}_j]=i\,\delta_{ij}$ and the harmonic excitations of
this system have the dispersion relation
$\Omega_k=Q\,\varepsilon_{{}_{\rm{K}}}\sqrt{1+4R^2\sin^2\frac{k}2}$.

We assume independent baths coupled to each degree of freedom of the
chain~\cite{CFTV99}, so that for this system Eq.~(\ref{e.FZPo}) is
easily generalized as
\begin{equation}
 {\cal Z}_P = C t^{PM\over 2}\prod_{i=1}^M\int d\bar
 x_i\int \prod_{n=1}^N da_{in}\,db_{in}~ e^{-S_P}~,
\label{Zp.phi4}
\end{equation}
where the action reads
\begin{eqnarray}
 S_P &=& \sum_{i=1}^M\Bigg\{\sum_{n=1}^N
 \bigg[ \frac{6tP^2}{Q^2R}~\sin^2\frac{\pi n}P +\frac3{2Rt}K_n \bigg]
 (a_{in}^2+b_{in}^2)
\nonumber\\ &&\hspace{12mm}
  + \frac{3R}{4t}(\bar q_i{-}\bar q_{i-1})^2+
\nonumber\\ &&\hspace{5mm}
 {+}\frac{3R}{2t}~\sum_{n=1}^N
 \Big[(a_{in}{-}a_{i-1,n})^2{+}(b_{in}{-}b_{i-1,n})^2\Big]
\nonumber\\ &&\hspace{22mm}
 +\frac{3}{2 R t P}~\sum_{\ell=1}^P v(q_{i\ell})\Bigg\}\,,
\end{eqnarray}
with the coordinates expressed in terms of their Fourier components as
\begin{equation}
 q_{i\ell}=\bar{q}_i+2\sum_{n=1}^N \Big(a_{in}\,\cos\frac{2\pi n\ell}P
 + b_{in}\,\sin\frac{2\pi n\ell}P \Big)~,
\end{equation}
and the dimensionless temperature reads
$t=(\beta\varepsilon_{{}_{\rm{K}}})^{-1}$.

The average quantities for the dissipative $\phi^4$ chain presented in
the figures have been obtained for periodic chains of length
($\sim{10^2}$ sites) large enough to be representative of the
thermodynamic limit for each set of physical parameters and by
extrapolating to $P\to\infty$ the results given by simulations at
finite $P$. A Drude-like spectral density, as introduced in
Section~\ref{ss.OhmDru}, was assumed for the environmental interaction,
so that the dissipative kernel reads
\begin{equation}
 K_n\equiv\frac{k_n}{\,\Omega^2}=\frac{\Gamma\Omega_{\rm{D}}}{1+Q\,\Omega_{\rm D}/(2\pi tn)}~,
\label{drudekern}
\end{equation}
where the dissipation strength $\Gamma\equiv\gamma/\Omega$ and the
cut-off frequency $\Omega_{\rm{D}}\equiv\omega_{\rm{D}}/\Omega$ are
also measured in units of the characteristic frequency
$\Omega=Q\varepsilon_{{}_{\rm{K}}}$.

\begin{figure}
\includegraphics[bbllx=14mm,bblly=91mm,bburx=175mm,bbury=208mm,
 width=80mm,angle=0]{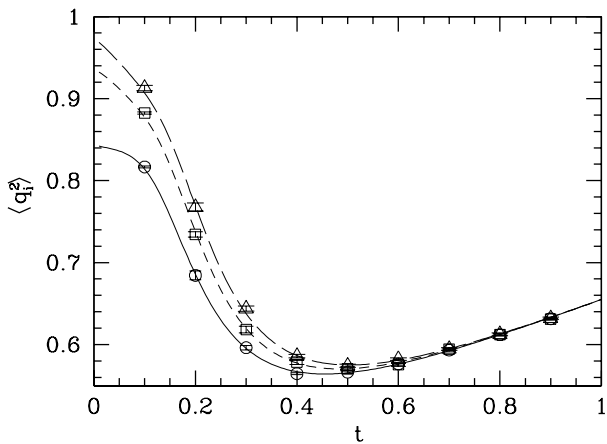}
\caption{\label{f.f4_1q2}
$\langle{{q}_i^2}\rangle$ vs temperature for the $\phi^4$ chain, with
$Q=0.2$, $R=5$, $\Omega_{\rm D}=100$ and different values of $\Gamma$.
The empty symbols are PIMC data (extrapolated for $P\to\infty$) and the
lines are the predictions from the effective potential
method~\cite{CFTV99}. $\Gamma=0$: circles and solid line; $\Gamma=20$:
squares and short-dashed line; $\Gamma=100$: triangles and long-dashed
line. The inset reports the average $\langle{v({q}_i)}\rangle$.}
\end{figure}

\begin{figure}
\includegraphics[bbllx=10mm,bblly=90mm,bburx=178mm,bbury=205mm,
 width=80mm,angle=0]{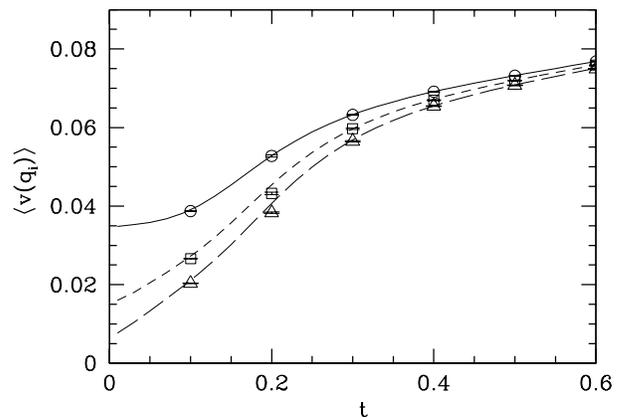}
\caption{\label{f.f4_1vq}
$\langle{v({q}_i)}\rangle$ vs temperature for the $\phi^4$ chain, with
$Q=0.2$, $R=5$, $\Omega_{\rm D}=100$ and different values of $\Gamma$.
Symbols and lines as in Fig.~\ref{f.f4_1q2}. }
\end{figure}

\begin{figure}
\includegraphics[bbllx=14mm,bblly=91mm,bburx=174mm,bbury=207mm,
 width=80mm,angle=0]{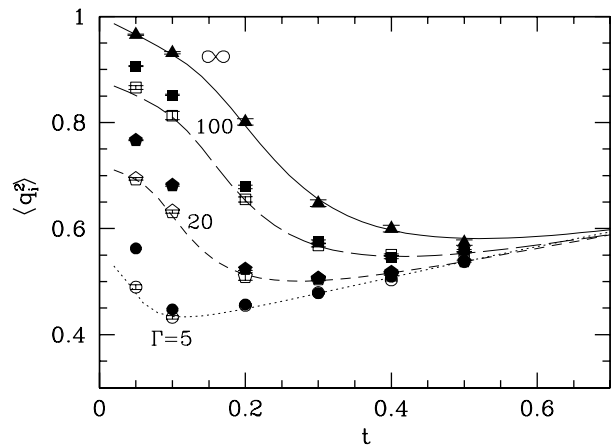}
\caption{\label{f.f4_2q2}
$\langle{q_i^2}\rangle$ vs temperature for the $\phi^4$ chain, with
$Q=1$, $R=3$, $\Omega_{\rm{D}}=10$ and different values of $\Gamma$.
The full symbols are PIMC data at finite Trotter number ($P=81$ for
finite $\Gamma$ and $P=11$ for $\Gamma\to\infty$); the empty symbols
are the extrapolated values for $P\to\infty$. The lines are the
predictions from the effective potential method~\cite{CFTV99}. }
\end{figure}

The comparison of our PIMC results with those of the
effective-potential method~\cite{CFTV99}, shown in Figs.~\ref{f.f4_1q2}
and~\ref{f.f4_1vq}, clearly indicates that the predictions of the
latter are very accurate; this accuracy is preserved for fairly large
values of the quantum coupling, close to the predicted limits of
applicability of the effective-potential approximation, as it appears
in Fig.~\ref{f.f4_2q2} which reports data for $Q=1$.

Moreover, in order to get a reliable thermodynamic limit, finite-size
effects have to be negligible, i.e., the number $M$ of sites must be
large enough. In this condition, reaching high Trotter numbers becomes
more and more computationally demanding and the extrapolation to
$P\to\infty$ problematic. However, such difficulty can sometimes be
overcome by means of a simple trick~\cite{CMPTV95} devised to improve
the bare Monte Carlo outcomes. According to Eqs.~(38) and (44) of
Ref.~\onlinecite{CMPTV95} any  finite-$P$ PIMC estimate $G(P)$ of a
given thermodynamic quantity $G$ can be corrected by the known error
affecting the same quantity for the corresponding SCHA system (of
course, including the dissipative action). This error, that can be
calculated in a simple way, is just the difference between the `exact'
($P\to\infty$) SCHA value, $G^{(h)}_{\rm HA}$ and the finite-$P$ SCHA
estimate $G^{(h)}_{\rm HA}(P)$. Note that any thermodynamic quantity of
interest for a quadratic action in presence of dissipation at finite
$P$ can be obtained starting from the density matrix given by Eq.~(A14)
of Ref.~\onlinecite{CFTV99} with $w=0$ and ${\mathbf{C}}$ and
$\bm\Lambda$ given by Eqs.~(36) and~(37) of the same reference, with
$\infty$ replaced by $N$ in the limits of the summations. We thus
correct the bare PIMC data $G(P)$ to the improved values
\begin{equation}
 G_{\rm HA}(P)=G(P)+\left[G^{(h)}_{\rm HA}-G^{(h)}_{\rm HA}(P)\right]~.
\end{equation}

\begin{figure}[t]
\includegraphics[bbllx=13mm,bblly=89mm,bburx=175mm,bbury=207mm,
 width=80mm,angle=0]{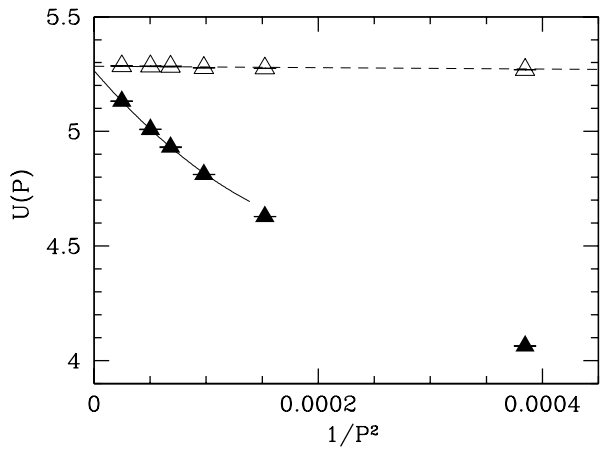}
\caption{\label{f.f4_2en}
Internal energy (per site) $U$ vs $1/P^2$ for the $\phi^4$ chain with
$Q=1$, $R=3$, $\Omega_{\rm{D}}=10$, $\Gamma=20$, at the temperature
$t=0.2$. The full triangles are the bare PIMC results $U(P)$, while the
empty ones report the harmonically-corrected data $U_{\rm{HA}}(P)$. The
lines are linear fits. }
\end{figure}

This procedure is shown to be very effective also for dissipative
systems, as shown in Fig.~\ref{f.f4_2en}, where the improved estimates
for the internal energy $U_{\rm{HA}}(P)$ display a very weak dependence
on $P$, at variance with the bare ones $U(P)$, allowing us to correctly
extrapolate to $P\to\infty$ by using much smaller values of $P$. The
relevance of the harmonic correction, which fully includes the
dissipation, increases with the dissipation strength.

We think that the above formulation of the Fourier path-integral Monte
Carlo can make it affordable to investigate the thermodynamics of
quantum many-body dissipative systems. The examples we reported testify
to the power of the method and confirm that the effective-potential
approach~\cite{CRTV97,CFTV99} is valid in the expected parameter range
(weak quantum coupling and/or strong dissipation). The further
developments involve the implementation of the Fourier PIMC procedure
beyond the limits of the effective-potential method. It is expected
that it will permit to study the behavior of strongly quantum systems
in presence of dissipation and thus open the possibility to approach
problems like the dissipative transition in Josephson-junction arrays.



\begin{thebibliography}{10}
\vspace{-4mm}
\bibitem{Weiss99}
 U. Weiss, {\em Quantum Dissipative Systems} (World Scientific,
 Singapore, 2nd edition, 1999).
\bibitem{TakahideEA2000}
 Y.~Takahide, R.~Yagi, A.~Kanda, Y.~Ootuka, and S.~I.~Kobayashi,
 Phys. Rev. Lett. {\bf 85}, 1974 (2000).
\bibitem{CRTV97}
 A.~Cuccoli, A.~Rossi, V.~Tognetti, and R.~Vaia,
 Phys. Rev. E {\bf 55}, 4849 (1997).
\bibitem{CFTV99}
 A. Cuccoli, A. Fubini, V. Tognetti, and R. Vaia,
 Phys. Rev. E {\bf 60}, 231 (1999).
\bibitem{CaldeiraL81}
 A.~O. Caldeira and A.~J. Leggett,
 Phys. Rev. Lett. {\bf 46}, 211 (1981).
\bibitem{CaldeiraL83a}
 A.~O. Caldeira and A.~J. Leggett,
 Ann. of Phys. {\bf 149}, 374 (1983).
\bibitem{TakahashiI1984}
 M. Takahashi and I. Imada,
 J. Phys. Soc. Jpn. {\bf 53}, 963 and 3765 (1984).
\bibitem{Doll84}
 J.~D. Doll,
 J.~Chem. Phys. {\bf 81}, 3536 (1984).
\bibitem{DollCF85}
 J.~D. Doll, R.~D. Coalson, and D.~L. Freeman,
 Phys. Rev. Lett. {\bf 55}, 1 (1985).
\bibitem{TopperT92}
 R.Q. Topper and D.~G. Trulhar,
 J.~Chem. Phys. {\bf 97}, 3647 (1992).
\bibitem{EleftheriouDCF99}
 M.~Eleftheriou, J.~D. Doll, E.~Curotto, and D.~L. Freeman,
 J.~Chem. Phys. {\bf 110}, 6657 (1999).
\bibitem{NeirottiFD00}
 J.~P. Neirotti, D.~L. Freeman, and J.~D. Doll,
 J.~Chem. Phys. {\bf 112}, 3990 (2000).
\bibitem{MielkeST00}
 S.~L. Mielke, J.~Srinavasan, and D.~G. Trulhar,
 J.~Chem. Phys. {\bf 112}, 8758 (2000).
\bibitem{LobaughV1992}
 J. Lobaugh and G.~A. Voth,
 J. Chem. Phys. {\bf 97}, 4205 (1992).
\bibitem{Feynman1972}
 R.P.~Feynman, {\em Statistical Mechanics} (Benjamin, Reading, MA, 1972).
\bibitem{GT85}
 R.~Giachetti and V.~Tognetti, Phys. Rev. Lett. {\bf 55}, 912 (1985).
\bibitem{GT86}
 R.~Giachetti and V.~Tognetti, Phys. Rev. B {\bf 33}, 7647 (1986).
\bibitem{FK86}
 R.P.~Feynman and H.~Kleinert, Phys. Rev. {\bf A34}, 5080 (1986).
\bibitem{GTV88fi4}
 R. Giachetti, V. Tognetti, and R. Vaia,
 Phys. Rev. A {\bf 38}, 1521 (1988); {\bf 38}, 1638 (1988).
\bibitem{SchneiderS80}
 T. Schneider and E. Stoll,
 Phys. Rev. B {\bf 22}, 5317 (1980).
\bibitem{MakiT79}
 H. Takayama and K. Maki,
 Phys. Rev. B {\bf 20}, 5009 (1979).
\bibitem{CMPTV95}
 A. Cuccoli, A. Macchi, G. Pedrolli, V. Tognetti, and R. Vaia,
 Phys. Rev. B {\bf 51}, 12369 (1995).
\end{thebibliography}
\end{document}